# Non-Equilibrium Thermodynamics Framework to Address the Glass Transition

Yikun Ren[1*], Feixiang Xu[2] and Ming Lin[1]

[1*]School of Materials Science and Engineering, Jiujiang University, Jiujiang, 332005, Jiangxi, China.
[2]School of Emergent Soft Matter, South China University of Technology, Guangzhou, 510641, Guangdong, China.

*Corresponding author(s). E-mail(s): 6260004@jju.edu.cn;
Contributing authors: mofxxu@mail.scut.edu.cn; linmingzny@163.com;

**Abstract**

When the center of fluctuations, i.e., the nonequilibrium eigenphase, undergoes transformation, there emerge critical parameters that demonstrate insensitivity to fluctuation perturbations and even independence from the molecular physical properties of the system, while exhibiting pronounced efficacy in governing phase transition dynamics.In the context of polymer glass transitions, Flory's conjecture (or $C1$ in the WLF equation) represents such a longstanding yet unresolved critical parameter. To address this issue, we replace entropy variation with a sequence of microstates and provide an analytically tractable statistical description of non-ergodicity. Our theory rigorously demonstrates that reaching the critical parameter is a necessary condition for non-equilibrium transitions to occur. Due to correlations between the eigenphase's entropy and energy in non-equilibrium systems, any system with a given intrinsic structure inherently possesses a critical parameter that represents the limiting deviation from equilibrium at any temperature. Flory's conjecture exemplifies this, with our new theoretical critical void ratio at the glass transition boundary calculated to be 2.6%, which closely matches experimental observations of 2.5%–2.6% over the past 50 years.

**Main:**

Some critical parameters in non-equilibrium transitions, such as Flory's void percentage in polymer glass transitions[1, 2] and the Reynolds number in turbulence[3], are fundamental to understanding these processes. These parameters not only define the phase diagrams of non-equilibrium systems but also serve as the foundation for deriving other physical quantities. For example, all linear polymers undergo a transition from a rubbery state to a glassy state when the void percentage (referred to as "free volume percentage" by Flory, although it has been shown to closely approximate the void percentage[4, 5]) decreases below a critical threshold, which remains invariant with respect to molecular structure and cooling rate[6]. Angell[7] emphasized the significance of the $C1$ parameter—another descriptor of the critical void percentage—in his paper "Why $C1$ = 16–17 in the WLF equation is physical—and the fragility of polymers." In his view, this parameter is crucial for understanding concepts like polymer fragility. However, to date, no theories have been able to predict the correct value of the critical void percentage, either before or after its experimental determination.

The experimental void percentage, also known as the $C1$ parameter in the Williams-Landel-Ferry (WLF) equation[6], is a widely used function for nearly all polymers and has attracted significant attention. Researchers have consistently attempted to analyze the $C1$ parameter using thermodynamic or kinetic equations based on equilibrium or near-equilibrium assumptions. However, only three theories are widely recognized as addressing the problem without the need for fitting: Simha-Boyer[8], Cohen-Grest[9], and Adam-Gibbs[10]. First, Simha-Boyer theory treats the space created by the intrinsic vibrations of molecules as free volume units, leading to a critical percentage approximately four times the experimental value. Second, Cohen-Grest theory defines free volume by identifying inflection points on the L-J potential energy curve, deriving a critical percentage roughly three times the experimental value. Finally, Adam-Gibbs theory suggests that the conformation of polymer chains is determined by cooperative chain movements and conformational entropy, resulting in groundbreaking findings. The critical percentage derived from this theory is the closest, nearly twice the experimental value. Despite these contributions, all three theories overlook a fundamental characteristic of critical parameter: it does not vary with molecular structure or cooling rate. The shortcomings of these theories stem from the assumptions underlying their formulation, highlighting the need for a new theoretical framework to properly explain these critical parameters.

Critical phenomena in statistical theory typically exhibit characteristics that are independent of molecular structure and the rates of physical processes. The novel theory is composed of three main components. First, we relax the assumption of local equilibrium, extending Prigogine's local concept into a more generalized framework. This revised concept satisfies the partition function law but does not inherently imply equilibrium. By exploring the union and intersection of multiple contiguous regions, we identify a critical scale, referred to as the Mean Area (MA), formed by different local regions under the influence of dynamical heterogeneity. This new mesoscopic scale possesses distinct properties, which will be discussed in detail later, and the method for determining its size will be explained in subsequent sections. In the second step, for the center of fluctuations，we demonstrate that the set of microstates within the MA corresponds to the truncation of a specific sequence, known as the Microstate Sequence (MSS), on the equipotential surface of the non-equilibrium systemIn a general context, the center of fluctuations in equilibrium systems constitutes a time-invariant eigenphase. In contrast, the nonequilibrium fluctuation center manifests as a dynamical entity whose temporal evolution is governed by fluctuation tendencies. When a nonequilibrium system becomes sufficiently removed from thermodynamic equilibrium, the influence of the fluctuation center on phase transition behavior assumes increasing significance. Hence, M-theory and fluctuation theory exhibit no fundamental incompatibility in their axiomatic foundations, but rather represent constitutively compatible and theoretically integrable frameworks. A detailed proof of this equivalence is provided in Supplementary Information A. Thus, the problem of minimizing non-equilibrium entropy is transformed into determining and truncating a directed sequence of microstates. This transformation can be addressed using the MSS theory developed in our former work in the context of the three-dimensional Ising model[11]. Finally, by inverting how microstate sequences contribute to the construction of MA, we define a key parameter, the r-parameter. This parameter quantifies the ensemble average of the thermodynamic probability of a particle being in a non-equilibrium state relative to its thermodynamic probability in equilibrium. Given its physical significance, a solution for the r-parameter can be derived within the framework of the dynamical mean-field method. By combining the entropy function from the second step with the solution for the r-parameter, we can analyze the critical parameter to better understand the physics behind non-equilibrium transitions.

In the following paragraphs, we develop a comprehensive theoretical framework,

referred to as M-theory, and provide two interrelated definitions for the degree of non-equilibrium. We then validate the corollaries of this framework using a polymer model, which leads to a novel intrinsic explanation of the polymer glass transition. By considering voids as the "solvent" and polymer chains as the "solute," we propose that the essence of the polymer glass transition lies in the attainment of the "supersaturation" limit. The core of our proof is that the theoretical value of the $C1$ parameter we derived closely matches the experimental value.

**Foundations of the M-Theory**

The partition function law states that the probability volume of any possible microstate in phase space is inversely related to its energy:配分定律

$$P_j = \frac{e^{-\frac{E_j}{kT}}}{z} \qquad (1)$$

$$z = \sum_j e^{-\frac{E_j}{kT}}$$

where $P_j$ is the probability of microstate $j$, and the partition function z runs over all microstates of the equilibrium system. This law is inherently derived from a fundamental physical viewpoint: there exists an intrinsic connection between thermodynamic probability and energy in phases with large number of particles.However, non-equilibrium systems do not satisfy this hypothesis due to fluctuations, dynamical heterogeneities, and macroscopic inhomogeneities. This does not imply a lack of applicability of thermodynamic laws to non-equilibrium systems; consequently, Prigogine proposed a mesoscopic approach to address this. He spatially divides non-equilibrium systems into numerous locals. Each local, being sufficiently small in volume, can be regarded as a "small equilibrium system". Different locals attain distinct equilibrium states under the influence of dynamic heterogeneities, thereby accounting for the non-ergodicity of the entire non-equilibrium system.

Building upon Prigogine's ideas, we have made two improvements. Firstly, we do not necessitate a strictly equilibrium local. This is because the dynamic relaxation patterns of many non-equilibrium systems render such a local non-existent. Hence, we propose a new local assumption:Local Assumption: Regardless of the type of system, there always exists a mesoscopic "locale" that is sufficiently small in macroscopic dimensions but large enough in microscopic dimensions to be considered a "point," with all its microscopic states conforming to the partition function law.

Nonetheless, for the locales (statistical units that are sufficiently small to be considered "points" but still contain a sufficient number of particles), the prerequisites for the validity of this hypothesis are inherently satisfied. Therefore, the partition function law is applicable to locales in systems where the local relaxation rate is sufficient to satisfy the local equilibrium hypothesis, such as polymers that adhere to the Onsager variational hypothesis.

Subsequently, we establish a new mesoscopic scale larger than the local scale, thereby transforming the approach of explaining non-ergodicity through dynamic heterogeneities into a statistical method for interpreting non-ergodicity.This larger mesoscopic scale is defined as follows: As we gradually zoom in from the macroscopic scale to the microscopic scale, there always exists a particular mesoscopic scale, denoted as the Mean Area (MA). When the observation scale is above the MA, the observer cannot detect dynamic heterogeneities but can observe the macroscopic statistical behavior of the system within a relatively short period of time. Conversely, when the observation scale is below the MA, the observer can detect dynamic heterogeneities but must observe over a longer period of time to grasp the macroscopic statistical behavior of the system. To provide a more rigorous mathematical and physical definition,, based on the Local Assumption, we coarse-grain both the MA and the locale into $n$ averaged particles (hereinafter referred to as "$n$-coarse-grained," where $n$ is sufficiently large). The intersection of the sets of microscopic states of all locals is always contained within the set of microscopic states of the Mean Area (MA), and simultaneously, the set of microscopic states of the MA is always contained within the union of the sets of microscopic states of all locals., According to the definition of Mean Area, the mathematical definition of the Mean Area can be stated as: "The set of microstates of the coarse-grained MA is equal to the intersection of the sets of microstates of all coarse-grained locales," as shown in Equation (3).

$$\{microstates\ in\ CG_{MA}\} = \bigcap_{locale \in MA} \{microstates\ in\ CG_{local}\} \qquad (3)$$

By combining Equation (2) with the Local Assumption, it can be inferred that the $n$-coarse-grained Mean Area formally satisfies the partition function law. Based on the second law of thermodynamics, the entropy of the Mean Area satisfies the following variational equation:

$$(\delta S)_{\overline{E_{total}}} = \left(\delta \sum_{local} S_{local}\right)_{\overline{E_{total}}} = \left(\sum_{local} \delta S_{local}\right)_{\overline{E_{total}}} = 0 \qquad (4)$$

Equation (4) implies that, under the constraint of the total energy of the non-equilibrium system, the overall entropy of the Mean Area reaches its maximum as each local region within the Mean Area attains maximum entropy.

It should be noted that the coarse-grained (CG) representation of the Mean Area does not enumerate all possible states, which is an inevitable consequence of superimposing local states in a non-equilibrium system. To identify the specific microstates present in the MA, we need to solve Equation (5). To facilitate this, we define an intermediate function that represents the sequence of newly emerging microstates within the Mean Area at any given moment, arising from various causes, as described in the Introduction. The second law of thermodynamics can be invoked to show that Equation (4) is equivalent to the monotonic variation of the MSS with respect to the system's energy. The proof is provided in Supplementary Information A. It is important to note that MSS is a descriptive framework for MA trajectories, not a complete representation of the system's true phase trajectories. MSS does not capture fluctuations or metastates specifically, but it can still reflect statistical regularities. Any microstate and its sequence index (usually a positive integer) on the MSS form a function $f$:

$$microstate \xrightarrow{f} i \qquad (5)$$

By combining Equations (4) and (5), we define the extent of non-equilibrium and special extent of non-equilibrium as:

$$\frac{dS}{di} \geq 0 \Rightarrow \{f^{-1}(microstates\ in\ MA)\} = \{1, 2, 3, \ldots, i_{max}\}$$

$$\omega = W/i_{max} \qquad (6)$$

$$r(\omega) = \frac{\sum_{i=1}^{W/\omega} e^{-E_i(s)/kT}}{\sum_{i=1}^{W} e^{-E_i(s)/kT}} \approx \frac{P(n)}{P(e)} \quad \#(7)$$

$W$ is the number of microstates of an equivalent system, and $i_{max}$ represents both the maximum sequence number of microstates and the total number of microstates in the non-equilibrium coarse-grained Mean Area. $\omega$ is the extent of non-equilibrium, which represents the degree of non-ergodicity of the whole system. $E_i(s)$ is the energy of a single particle s in microstate i. r is the special extent of non-equilibrium, which represents the degree of non-ergodicity of a single particle. P(n) is the probability of particle s choosing to be "non-equilibrium states". And P(e) is the probability of particle s choosing to be "equilibrium states".

The function of the MSS index in the Ising model[11] can be written as:

$$i = f^{-1}(microstates\ in\ MA) = \begin{cases} 1 + \sum_{j=0}^{\rho-1} C_N^j + \sum_{k=1}^{\rho} h(k), \rho > 0 \\ 1, \rho = 0 \end{cases} \quad (8)$$

The special extent of non-equilibrium described by Equation (7) is useful but complicated. To address the exact relationship of r and $\omega$, we consider the thermodynamic probability of an arbitrary microstate within the system, denoted as $f(i_p, N) = P(i)$, where $i_p$ is the number of particles in a non-equilibrium state and $N$ is the total number of particles, i is the microstate sequence index. For a given local region, if we neglect the impact of dynamical heterogeneity on boundary conditions, the partition function can be expressed as:

$$Z_{local} = Z_{equilibrium} = \sum_{ip=0}^{N} f(i_p, N) = \sum_{i}^{W} P(i) \quad (9)$$

Now, let us consider the partition function for a Mean Area with equation (7) and (9):

$$Z_{MA} = \sum_{i}^{W/\omega} P(i) = \frac{\sum_{ip=0}^{N} f(i_p, N)}{\omega r} \quad (10)$$

Equations (9) and (10) describe the transition of a coarse-grained microscopic particle from a state unaffected by the degree of non-equilibrium to one influenced by it. Based on the definition of r, there is an equivalent expression for the partition function of the Mean Area (MA):

$$Z_{MA} = \sum_{ip=0}^{N} f(i_p * r, N) \quad \#(10)$$

Thus, there exists an identity:

$$\lim_{N \to \infty} \frac{\sum_{ip=0}^{N} f(i_p, N)}{\omega r} - \sum_{ip=0}^{N} f(i_p * r, N) = 0 \quad (11)$$

The $r$ parameter that satisfies Equation (11) is the microscopic representation of the degree of non-equilibrium. It can be interpreted as the probability of each spin selecting a state closer to non-equilibrium (note that when the system reaches equilibrium, the probabilities for the spin to take different values are equal). This implies that if the structure of the non-equilibrium system is known, the $r$ parameter can be precisely calculated from a microscopic perspective. Furthermore, $r$ is also related to the extent of non-equilibrium, which can be solved using the MSS approach. Therefore, the critical parameters for the non-equilibrium transition can be accurately determined by calculating the $r$ parameter.

$$\omega r = \frac{Z_{local}}{Z_{MA}} \quad (12)$$

As defined in Equation (12) and (7), the parameter $r$ is a correction to the thermodynamic probability for each particle. It is related to the non-equilibrium dynamics of individual particles, thus having broader research significance.

After the above derivation, we have derived complete expressions for the extent and special extent of non-equilibrium, microstate sequence and microstate sequence index based on the Mean Area and Microstate Sequence framework. The new theory, incorporating MA and MSS, can be termed M-theory. Therefore, we can derive the following two related equations:

First, as the degree of non-equilibrium approaches 1, the system approaches equilibrium. According to the partition function principle and Equation (4), there exists a reciprocal relationship between non-equilibrium entropy and the degree of non-equilibrium, which can be expressed as:

$$S \sim S\left(\frac{1}{\omega}\right) \rightarrow S = \frac{a_1}{\omega} + \frac{a_2}{\omega^2} \dots \quad (13)$$

Second, the self-driving force for a non-equilibrium system to approach equilibrium originates from the non-equilibrium free energy. This driving force, $f$ is positively correlated with the extent of non-equilibrium, which can be expressed as:

$$E = \frac{\sum_{i=1}^{w/\omega} E_i e^{-E_i/kT}}{\sum_{i=1}^{w/\omega} e^{-E_i/kT}} = e_0 + \frac{e_1}{\omega} \dots$$
$$f = f_1(\omega - 1) + f_2(\omega - 1)^2 \dots$$
$$E - ST \geq fT \quad (14)$$

We take the constant term and the first term of the Taylor expansion mentioned above to perform a basic analysis of the situation represented by the equality in Equation (14):

$$e_0 + (e_1 - a_1 T)/\omega \geq f_1(\omega - 1)T$$
$$\rightarrow \omega \leq \omega_{max}(T) \quad (15)$$

Equation (13) has many possible solutions. However, in this paper, we will focus on the implications of the maximum value of the degree of non-equilibrium. By combining Equations (13) and (14), we can derive the following equivalent conclusions:

1. The extent of non-equilibrium of any system at a given temperature is limited. That is, without altering the system's structure, its complexity remains finite.

2. The non-equilibrium entropy of any system at a given temperature is also limited.

3. The upper limit of the extent of non-equilibrium forms a temperature-dependent function:

$$C = M(E - ST - f(\omega)T) \quad (16)$$

And this value is always constant. In other words, there exists a critical parameter that divides non-equilibrium phases and determines the conditions under which non-equilibrium phase transitions occur.

Next, by leveraging the correlation between the Ising model and the polymer lattice model, we will rigorously prove all the conclusions of M-theory within the context of the polymer lattice.

**Analysis of the Critical Void Percentage in Polymer Glass Transition**

Although the key concept in M-theory—the extent of non-equilibrium—is a physical concept that transcends specific mathematical forms, to rigorously demonstrate the validity of the new theory in real-world transformation behaviors, we still derive it starting from the Ising model. This approach is valuable for verifying the theoretical framework. However, the polymer system cannot be conveniently transformed into an Ising model in the same way as the lattice gas model. It requires a projection technique to simplify the polymer lattice model into a tractable Ising model form with intrinsic constraints:

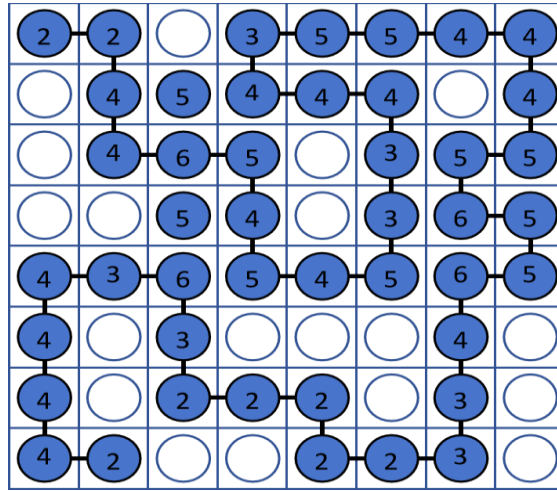

Figure 1: Simplified Simple Cubic Diagram of the Polymer Lattice Model

As shown in Fig. 1, the polymer can be simplified into a lattice model. The blue balls represent segments, and the white balls represent vacancies. It is important to clarify that these white balls represent vacancies, not free volume. In other words, our analysis focuses strictly on the proportion of vacancies within a spatial context. The total volume of the system can then be expressed as:

$$V_{total} = V_{vacancy} + V_b \cdot gL \tag{17}$$

where $V_{total}$ is the total volume, $V_{vacancy}$ is the vacancy volume, $V_b$ is the volume of a single lattice site, $g$ is the number of chains, and $L$ is the number of segments in a single chain.

Note that we need to convert the three-dimensional conformation of the polymer into a one-dimensional Ising-like sequence. To do this, we assign a number to each segment to record the total number of nearest neighbors. We then ignore the chain endpoints and classify the spins into two categories: spins with a value of 6, and spins that can take values between 2 and 5. Spins with a value of 1 correspond to chain endpoints, which can be ignored due to the assumption that the polymer chains are very long.

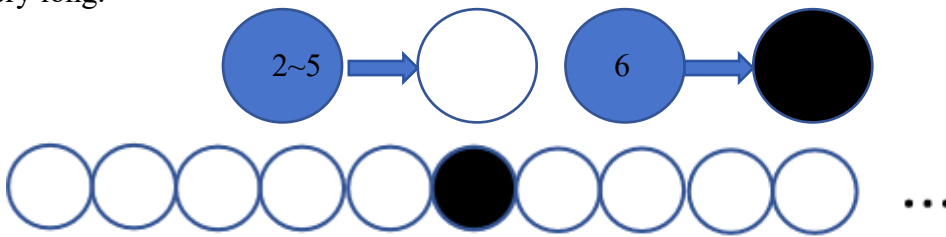

Figure 2. Simplification of Spin Values: Spins of 2-5 are treated as a single type of white spin, while spins of 6 are treated as a different type.

The specific transformation is shown in Figure 2. Next, using the contact relationship between the voids and the segments on the lattice, we deduce that:

$$\frac{V_{vacancy}}{V_{total}} \approx \frac{n_{black}}{n_{total}} \tag{18}$$

Following the general notation of MSS theory, $\rho$ represents the number of spins with a value of 6, and $n$ is the total number of segment spins. The relaxation process involves the reduction of vacancies. Therefore, the direction of incremental increase for spins with a value of 6 indicates the relaxation direction. A spatially constrained Ising model can be deduced from the Ising model formulation as follows:

$$i = \begin{cases} 1 + \sum_{j=0}^{\rho-1} C_N^j + \sum_{k=1}^{\rho} h(k), & \rho > 0 \\ 1, & \rho = 0 \end{cases} \tag{19}$$

The energy and extent of non-equilibrium in the polymer lattice system are:

$$E \approx -\frac{\sum_{\rho=0}^{\rho(i)}(J_1\rho + J_2(N-\rho))C_N^\rho}{\sum_{\rho=0}^{\rho(i)} C_N^\rho} \tag{20}$$

$$\omega = 2^N / i_{max} \tag{21}$$

Based on the definitions of the extent of non-equilibrium and non-equilibrium entropy, we derive an expression for the polymer entropy and analyze its minimum point. First, we decompose the set of non-equilibrium microscopic states into more manageable subsets. Following this approach, each subset represents a scenario where only $m$ segments are in a non-equilibrium state, while the remaining segments are all in a lower-energy state. As $m$ increases, more segments participate in the formation of the non-equilibrium state, until the current non-equilibrium system is achieved. This mathematically translates the system's non-equilibrium properties to each individual segment, enabling us to analyze phase transition behavior by combining the local characteristics of the segments after solving the entropy function. According to Equations (11), (12) and (18), we obtain:

$$\frac{2^L}{\omega r} = \sum_{m=0\sim L} W(m,\omega) = C_L^L + C_L^{L-r} + C_L^{L-2r} + C_L^{L-3r} + \cdots + C_L^{L-Lr} \tag{22}$$

$$r(\omega) \approx \frac{P(n)}{P(e)} \approx \frac{P(segment)}{P(void)} \tag{23}$$

If the system reaches equilibrium, $r=1$. Directly applying these equations is challenging, prompting us to use appropriate mathematical approximations. Next, we represent the probability of each microstate as the product of multiple conditional probabilities, where each conditional probability corresponds to the probability of each black ball. Consequently, we obtain the following expression:

$$P_j ln P_j = \prod_{m=1}^{\rho} f(m,j) \tag{24}$$

where $j$ is the microstate sequence index, $f(m, j)$ is the intermediate function, and $m$ is the index number of the black ball.

$$S(\omega) = -k \sum_{j=1}^{w/\omega} P_j ln P_j = -kN \sum_{j=1}^{w/\omega} \prod_{m=1}^{L} f(m,j)$$

$$= -kN \prod_m \sum_j f(m,j) = -kN \prod_m \overline{f(\omega)} \tag{25}$$

By swapping the order of multiplication and addition, we express the conditional

probability as the average conditional probability over all microstates. From Equation (23), we obtain:

$$\overline{f(\omega)} = \sum_j \frac{P_j(m)lnP_j(m)}{P_j(m-1)lnP_j(m-1)}$$

$$= \sum_j \frac{lne^{-\beta E_j(m)}/z(m)}{lne^{-\beta E_j(m-1)}/z(m-1)} \frac{e^{-\beta E_j(m)}/z(m)}{e^{-\beta E_j(m-1)}/z(m-1)}$$

$$\approx C_N^m \frac{lne^{-\beta E_j(\rho)}/z(m)}{lne^{-\beta E_j(\rho-1)}/z(m-1)} \frac{e^{-\beta E_j(\rho)}/z(m)}{e^{-\beta E_j(\rho-1)}/z(m-1)}$$

$$= C_N^m (1 + \frac{\beta \Delta E(\rho(\omega))}{lnW}) \frac{z(m)}{z(m-1)}$$

$$= C_N^m \left(1 + \frac{\beta \Delta E(\rho(\omega))}{lnW}\right) e^{-\frac{\Delta E}{kT}} \frac{c_L^{L-mr}}{c_L^{L-(m-1)r}} \quad (26)$$

Here, N is the total number of segments, and $\Delta E = -\sum_{\rho=0}^{\rho(i)}(J_1-J_2)\rho C_N^\rho/2^N = \Delta E(\omega)$. The parameter $r$ is defined in equations (22) and (23).

Entropy should always increase or remain constant. However, the presence of parameter $r$ inevitably reduces the rate of entropy increase. This causes segments with larger spin values to aggregate, slowing the process of approaching equilibrium. The question is whether there exists a critical value of $r$ that completely halts the relaxation process and prevents further supercooling. Let us analyze the entropy function (note that no fitting parameters have been used in the derivations).

$$Q = N\lambda C_N^{\rho(\omega)} \left(1 + \frac{\beta \Delta E(\rho(\omega))}{lnAW}\right) \quad (27)$$

$$S(\omega) = -kQ \prod_{m=1}^{GL}(1 - \frac{r}{GL-m})^L$$

$$= -kQ \prod_{g=0}^{G} \frac{1}{N-L}^L \prod_{g=0}^{G}(N-gLr)^L = -kQD \prod_{g=0}^{G}(N-gLr)^L \quad (28)$$

$$D = \prod_{\rho=0}^{\rho(\omega)} \left(\frac{1}{N-L}\right)^L \quad (29)$$

Then,

$$[(N-Lr)^{rL}(N-2Lr)^{rL}(N-3Lr)^{rL}\ldots(N-gLr)^{rL}]^{1/r}$$

$$\approx \left[\frac{N!}{(N-rL)!}\frac{(N-rL)!}{(N-2rL)!}\ldots\frac{(N-(G-1)Lr)!}{(N-GLr)!}\right]^{\frac{1}{r}} = \left[\frac{N!}{(N-GLr)!}\right]^{\frac{1}{r}} \quad (30)$$

Substituting the intermediate equation (30) into equation (28):

$$S = -kQD\left(\frac{N!}{(N-GLr)!}\right)^{\frac{1}{r}} \quad (31)$$

A careful examination of Equation (31) reveals that non-equilibrium primarily affects the entropy function of polymers through the parameter $r$. Note that the critical point corresponding to the divergence of the partial derivative of entropy is given by the following equation:

$$N - GLr = 0 \quad (32)$$

It is important to note that Equation (32) represents the scenario where the spatial utilization rate of the system has reached its limit. This point is both a mathematical critical point and a physical extremum. It demonstrates the existence of a limit to the degree of non-equilibrium, which is manifested through critical parameters. By employing mathematical methods to eliminate the effect of chain ends on the entropy[12], we can modify the critical equation as follows:

$$N - GLr = 0 \rightarrow 1 - \frac{r\left(\rho - \frac{2(\rho-1)}{a}\right)}{N - \frac{2(\rho-1)}{a}} = 0 \rightarrow \frac{3}{1-\emptyset} = 2r(\omega) + 1 \quad (33)$$

$$\emptyset = \frac{V_{vacancy}}{V_{total}} = \frac{N - gL}{N} \approx \frac{n - \rho(\omega)}{n} \quad (34)$$

Thus, we obtain a function that depends only on the extent of non-equilibrium $\omega$ and temperature:

$$\frac{3n}{\rho(\omega)} = 2e^{\frac{\Delta E(\omega)}{kT}} + 1 \quad (35)$$

It should be noted that this critical equation is a definitive equation for the extent of non-equilibrium, denoted by $\omega$. However, Equation (33) is overly complex. Its equivalent form, Equation (31), is more suitable for exact solution due to the property of $r$ in Equation (22) and (23).

Figure 3 illustrates a treatment methodology called the micro-scissor ideal experiment: Imagine using extremely tiny scissors to cut the links between segment sites while preserving the energy or state of each site. In this way, while the energy, non-equilibrium extent, and interactions between segment sites remain unchanged, the degrees of freedom are altered. Consequently, the segment sites expand. Concurrently, energy is carefully absorbed or released to the system from the external environment, ensuring that the average kinetic energy per site remains constant. The average free path of the segment sites after cutting is denoted as $l$. The cube of this free path, and the cube of the average free path between vacancy sites (i.e., the size of vacancy sites, $b$), can be used to express the parameter $r$:

$$r = \frac{P_{(segment)}}{P_{(void)}} = \frac{\frac{1}{l^3}}{\frac{1}{b^3}} = \frac{b^3}{l^3} \quad (36)$$

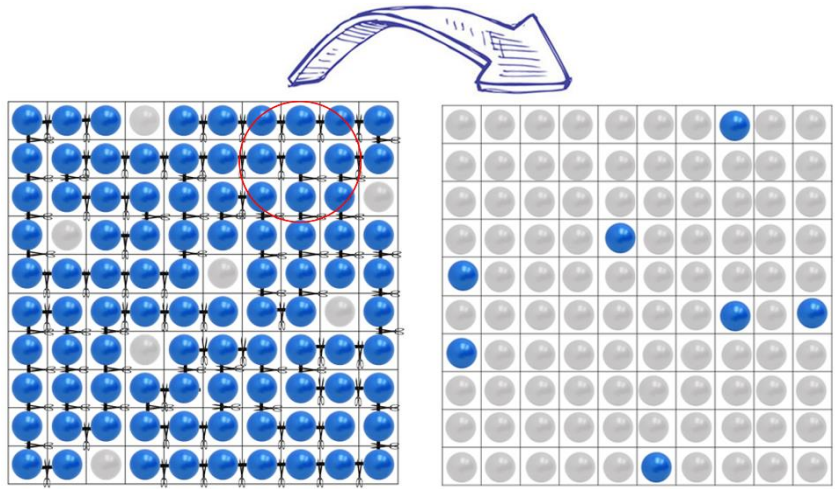

Figure 3. Polymer lattice after cutting in the "micro-scissor experiment". Left: before cutting; right: after cutting.

Let us derive a new expression for the parameter $r$ at the glass transition point. The vacancies in the system can be considered equivalent to the 'vacancies' created by the movement of segment sites. Treating the entire lattice as a confined system, the increased mobility of segment sites causes the vacancies to expand by a factor of 36 after 'cutting'. This expansion results from a six-fold increase in the motion modes and a six-fold shift in the center of mass. Thus:

$$r = \frac{b^3}{l^3} = \frac{(N-gL)b^3}{(N-gL)l^3} = \frac{(N-gL)b^3}{36gLb^3} \qquad (37)$$

By combining Equation (33) and Equation (37), we obtain:

$$\frac{3}{1-\emptyset} = 2\frac{1-\emptyset}{36\emptyset} + 1 \qquad (38)$$

The final result is expressed as:

$$\emptyset = \frac{-19 + 3\sqrt{42}}{17} \approx 2.6\% \qquad (39)$$

Converted to the $C1$ parameter, it is 1/2.303*2.6%=16.7, representing the smallest experimental deviation among all theories, as shown in the comparison in Table 1. I have demonstrated the consistency between the theoretical and experimental values of the invariant through rigorous mathematical analysis, without introducing any fitting processes or unknown parameters. Therefore, our theory proposes that the glass transition is a quasi-phase transition with a critical constant represents its limit of degree of non-equilibrium at any temperature point. The exact value of our theory and other non-fitting theories and experimental results have been shown in the table 1: .

| Methods | SimhaBoyer (1950) | Cohen (1970) | AdamGibbs (1980) | MSS result (Today's work) | Experiments (Current study) |
|---|---|---|---|---|---|
| Vacancy(%) | 11.6% | 7.6% | 5.3% | 2.6% | 2.5%~2.6% |
| $C1$ | 3.7 | 5.7 | 8.2 | 16.7 | 16.7~17.4 |

Table 1. Theoretical and experimental values of $C1$

Lastly, this constant has three physical meanings: First, it represents the point at which the non-equilibrium entropy or the degree of non-equilibrium reaches its limit at a given temperature; second, it signifies the critical boudary between the rubbery (or liquid) state and the glassy state; and third, it indicates that, in the absence of external forces, all the free energy of the non-equilibrium system is exhausted in driving its own structure to approach the equilibrium. A simpler way to understand glass transition is to consider voids as the "solvent" and polymer chains as the "solute." The critical point of the glass transition then represents the upper limit of the "supersaturation" of the polymer chains.

**Conclusion**

We build M- theory to study the transition caused by change of the center of fluctuations. The M-theory, based on Prigogine's concept, provides a solution to Flory's conjecture and demonstrates that the necessary condition for a non-equilibrium transition is reaching a critical parameter that represents the limiting value of the degree of non-equilibrium at any temperature and in any physical process. The Mean Area introduces a new critical mesoscopic perspective, while the Microstate Sequence serves as a statistical tool within this novel framework. This theory offers an intuitive understanding of non-ergodicity in the Ising model and proposes an explanation for non-equilibrium transitions caused by the transition of the center of fluctuations. Each non-equilibrium transition is associated with a specific critical parameter, with Flory's void percentage being one such example.

The MSS theory provides a rigorous proof for the glass transition. The glass transition in general linear polymers is identified as a second-order quasi-phase transition. The $C1$ parameter in the WLF equation is the only exact partitioning factor, free from fluctuations, that clearly marks the transition boundary. The value of $C1$, related to the vacancy volume issue, has been determined without any fitting and is found to be 16.7. For the first time, the theoretical and experimental values of $C1$ (16–17) agree within the experimental error margins.

Whether considering phase transitions in equilibrium systems or transformations between non-equilibrium states, non-equilibrium processes are essential. This theoretical framework provides a viable approach for studying non-equilibrium transitions under more conventional conditions, rather than being limited to the idealized scenario of the slowest processes. Therefore, it holds significant potential for analyzing a wide range of non-equilibrium transitions.

**Acknowledge**

This work was supported by the National Natural Science Foundation of China (grant number 21973033), the Fundamental Research Funds for Central Universities (grant number 2018ZD13), the Natural Science Foundation of Jiujiang Science and Technology Bureau (grant number S2022KXJJ001), and the China Postdoctoral Science Foundation (grant number 2024M753321).